\documentclass[a4paper,twoside,12pt]{article}
\usepackage{amstex}
\usepackage{a4}

\newdimen\control
\newdimen\target
\newdimen\laenge
\newdimen\increment
\def\CNOT(#1,#2,#3){%
\increment20sp
\control#3\increment
\target\control
\begin{picture}(30,\number\target)
\advance\control-#1\increment
\advance\target-#2\increment
\multiput(0,0)(0,20){#3}{\line(1,0){30}}
\put(15,\number\target){\circle{10}}
\put(15,\number\control){\makebox(0,0){$\bullet$}}
\ifnum #1 < #2
\laenge\control
\advance\laenge-\target
\advance\laenge5sp
\put(15,\number\control){\line(0,-1){\number\laenge}}
\else
\laenge\target
\advance\laenge-\control
\advance\laenge5sp
\put(15,\number\control){\line(0,1){\number\laenge}}
\fi
\end{picture}\ignorespaces}

\newtheorem{theorem}{Theorem}[section] 
\newtheorem{lemma}[theorem]{Lemma}

\newtheorem{algorithm}[theorem]{Algorithm}

\newcommand{\dirsum}[0]{\oplus}
\newcommand{\bigdirsum}[0]{\bigoplus}
\newcommand{\tensor}[0]{\otimes}
\newcommand{\ext}[0]{\overline}
\newcommand{\normal}[0]{\unlhd}
\newcommand{\DFT}[0]{\mbox{\rm DFT}}
\newcommand{\diag}[0]{\mbox{\rm diag}}
\newcommand{\ind}[0]{\uparrow}
\newcommand{\onemat}[0]{{\bf 1}}
\newcommand{\C}[0]{\mathbb{C}}
\newcommand{\qed}[0]{\hfill$\Box$}

\begin{document}

%
%

\begin{titlepage}

\begin{center}
{\Large Fast Quantum Fourier Transforms\\
for a Class of Non-abelian Groups\par}

\vspace*{1cm}
\begin{center}
\begin{tabular}{c@{\qquad}c}
Markus P\"uschel & Martin R\"otteler\footnotemark[1]\\
{\tt pueschel\symbol{64}ira.uka.de} & 
{\tt roettele\symbol{64}ira.uka.de}\\ 
\\
\multicolumn{2}{c}{Thomas Beth} \\ 
\multicolumn{2}{c}{\tt EISS\_Office\symbol{64}ira.uka.de}
\end{tabular}
\end{center}
\footnotetext[1]{supported by DFG grant GRK 209/3-98}

\vspace*{1cm}
Institut f\"ur Algorithmen und Kognitive Systeme\\
Universit\"at Karlsruhe, Germany

\vfill
{\bf Abstract}

\bigskip
\begin{minipage}{12cm}
\noindent
An algorithm is presented allowing the construction of fast Fourier
transforms for any solvable group on a classical computer. The special
structure of the recursion formula being the core of this algorithm
makes it a good starting point to obtain systematically fast Fourier
transforms for solvable groups on a quantum computer. The inherent
structure of the Hilbert space imposed by the qubit architecture
suggests to consider groups of order $2^n$ first (where $n$ is the
number of qubits). As an example, fast quantum Fourier transforms for
all 4 classes of non-abelian $2$-groups with cyclic normal subgroup of
index 2 are explicitly constructed in terms of quantum circuits. The
(quantum) complexity of the Fourier transform for these groups of size
$2^n$ is $O(n^2)$ in all cases.
\end{minipage}

\vfill\vfill
\end{center}
\end{titlepage}

%
%

\section{Introduction}

Quantum algorithms are a recent subject and of possibly central
importance in physics and computer science.  It has been shown that
there are problems on which a putative quantum computer could
outperform every classical computer. A striking example is Shor's
factoring algorithm (see \cite{Shor:94}).

Here we adress a problem used as a subroutine in almost all known
quantum algorithms: The quantum Fourier transform (QFT) and its
generalization to arbitrary finite groups.

In classical computation there exist elaborate methods for the
construction of Fourier transforms (see Beth \cite{Beth:84},
\cite{Beth:87}, Clausen \cite{Clausen:89}, \cite{Clausen:93}, and
Diaconis/Rockmore \cite{Diaconis:90}) so it is highly interesting to
adapt and modify these methods to get a quantum algorithm with a much
better performance (with respect to the common quantum complexity
model) as on a classical computer.

First attempts in this direction have been proposed by Beals
\cite{Beals:97} and H\o yer \cite{Hoyer:97}. In this paper we present
an algebraic approach using representation theory which can be seen as
a first step towards the realization of a large class of QFTs on a 
quantum computer.

%
%

\section{Generalized Fourier Transforms}\label{GFT}

Fourier transforms for finite groups are an interesting and well
studied topic for classical computers. We refer to \cite{Beth:84},
\cite{Clausen:93}, \cite{Maslen:95}, \cite{Pueschel:98} as
representatives for a vast number of publications. The reader not
familiar with the standard notation concerning group representations
should refer to these publications or to standard references as
\cite{Curtis:81} or \cite{Serre:77}.
 
For the convenience of the reader we first recall the definition of
the generalized Fourier transforms for a finite group $G$ and explain
the representation theoretical point of view we are going to take.

Each isomorphism 
\[
\Phi : \C G \longrightarrow \bigoplus_{i=1}^k \C^{d_i \times d_i}
\]
between the group algebra of $G$ and the Wedderburn components is
called a {\em Fourier transform} for the group $G$.  A particular
isomorphism is fixed by picking a system $\rho_1, \ldots, \rho_k$ of
representatives of irreducible representations of $G$ and defining
$\Phi$ as the linear extension of the mapping $g\mapsto
\bigdirsum_{i=1}^k \rho_i(g),\ g\in G$ (of course $\deg(\rho_i) =
d_i$). Any ``fast'' algorithm for the evaluation of $\Phi$ is called a
{\em fast Fourier transform} for $G$.

In order to obtain a matrix representation for the linear mapping
$\Phi$ one usually fixes {\em natural} bases $L$ in $\C G$ and
$L'$ in $\bigdirsum_{i=1}^k\C^{d_i \times d_i}$. This is done by choosing an
ordering $L=(g_1,\ldots, g_{|G|})$ on the elements of $G$ and an
ordering $L'$ on the elementary matrices $E_{k,l}$ (1 at position
$(k,l)$, 0 else) which correspond to the coefficients of the
irreducible representations appearing in the Wedderburn decomposition.
The Fourier transform $\Phi$ then is represented by a matrix $M_{L,L'}
\in \C^{|G|\times |G|}$. Since $M_{L, L'}$ is a base change between
two orthonormal bases (with respect to the standard hermitean scalar
product on $\C G$) it must be unitary.

The matrix $M_{L, L'}$ can also be characterized by another property.
Let $\phi$ be the regular representation obtained by right
multiplication of $G$ on $L$.  Conjugating $\phi$ with $M_{L, L'}$
yields (up to a permutation matrix $P$) a direct sum of irreducible
representations with the property that equivalent irreducibles 
are equal, i.e.  
$$
\phi^{M_{L, L'}} = \left(\rho_1\dirsum\dots\dirsum\rho_k\right)^P
\quad\mbox{fulfilling}\quad
\rho_i\cong\rho_j\Rightarrow\rho_i = \rho_j.
$$
On the other hand every matrix with this property corresponds to a
Fourier transform with respect to natural bases.  

As an example let $G = Z_n = \langle x \mid x^n = 1\rangle$ be the
cyclic group of order $n$ with irreducible representations $\rho_i =
(x\mapsto\omega_n^i)$, $i = 0\dots n-1$, where $\omega_n$ denotes a
primitive $n$-th root of unity.  With respect to the natural bases $L
= (x^i\mid i = 0\dots n-1)$ and $L' = (E_{1,1},\dots,E_{n,n})$ the
matrix $M_{L, L'} = \frac{1}{\sqrt{n}}\left[\omega_n^{i\cdot j}\mid
i,j = 0\dots n-1\right] = \DFT_n$ is the discrete Fourier
transform well-known from signal processing.

We will refer to the notion of a fast Fourier transform as a fast
algorithm for the multiplication with $M_{L, L'}$.  Of course, the
term ``fast'' depends on the chosen complexity model. Since we are
primarily interested in the realization of a fast Fourier transform on
a quantum computer (QFT) we first have to define the measure of
complexity on this architecture.

%
%

\section{The Complexity Model}\label{complexity}

In this paper we think of a quantum computer to consist of a quantum
register which in turn consists of $n$ qubits each of which provides a
$2$-dimensional Hilbert space. Thus the possible operations this computer
can perform are given by the unitary group ${\cal U}(2^n)$.

To study the complexity of unitary operators on $n$-qubit quantum
systems we introduce the following two types of building blocks:

\noindent
\begin{itemize}
\item Local unitary operations on the qubit $i$ are of the form
\[U^{(i)} := \onemat_{2^{i-1}} \otimes U \otimes \onemat_{2^{n-i}},
\]
where $U$ is an element of the unitary group ${\cal U}(2)$ of $2\times
2$-matrices. 
\item The controlled NOT gate (also called measurement gate)
between the qubits $i$ (control) and $j$ (target) is defined
by
\[
{\rm CNOT}^{(i, j)} := 
\left(
\renewcommand{\arraystretch}{0.2}%
\setlength{\arraycolsep}{0.2em}%
\newcommand{\n}{\ }
\begin{array}{cccc} 
1 & \n  & \n  & \n  \\
\n  & 1 & \n  & \n  \\
\n  & \n  & \n  & 1 \\
\n  & \n  & 1 & \n
\end{array}
\right)
\]
when restricted to the tensor component of the Hilbert space 
which is spanned by the qubits $i$ and $j$. Note that the
controlled not is nothing but an XOR on the resp. components.
\end{itemize}

\begin{figure}
\input{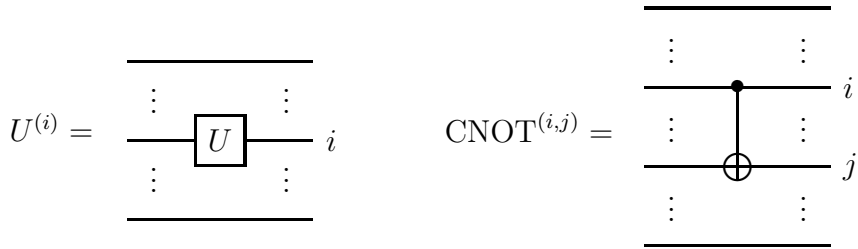}
\caption{\label{gates}Elementary quantum gates}
\end{figure}

\noindent
In the graphical notation using quantum wires these transforms are
written as shown in figure \ref{gates}. The lines represent the qubits
and the least significant bit is the lowest.
We assume that these so-called elementary quantum gates can be 
performed with cost $O(1)$. 

These two types of gates suffice to generate all unitary
transformations, which is the content of the following theorem 
from \cite{Barenco:95}.

\begin{theorem}
The set ${\cal G} = \{ U^{(i)}, {\rm CNOT}^{(i, j)}\mid
U\in{\cal U}(2),\ i,j=1\dots n,\ i\neq j \}$ is a generating set 
for the unitary group ${\cal U}(2^n)$. 
\end{theorem}

\noindent
This means that for each $U \in {\cal U}(2^n)$ there is a word $w_1
w_2 \ldots w_k$ (where $w_i \in {\cal G}$ for $i=1\dots k$ is an
elementary gate) such that $U$ factorizes as $U = w_1 w_2 \ldots w_k$.

In general only exponential upper bounds for the minimal length
occuring in factorizations have been proved (see \cite{Barenco:95})
but there are many interesting classes of unitary matrices in ${\cal
U}(2^n)$ affording only polylogarithmic word length, which means, that
the minimal length $k$ is asymptotically $O(p(n))$ where $p$ is a
polynomial.

In the following we give examples of some particular unitary
transforms admitting short factorizations which will be useful 
in the rest of the paper.

\begin{itemize}
\item
The symmetric group $S_n$ is embedded in ${\cal U}(2^n)$ by the
natural operation of $S_n$ on the tensor components (qubits).  Let
$\tau\in S_n$ and $\Pi_\tau$ the corresponding permutation matrix on
$2^n$ points. Then $\Pi_\tau$ has a $O(n)$ factorization as shown in
\cite{Moore:98}. As an example in figure \ref{123} the permutation
$(1,3,2)$ of the qubits (which corresponds to the permutation
$(2,5,3)(4,6,7)$ on the register) is factored according to $(1,3,2) =
(1,2)(2,3)$.

\begin{figure}
\input{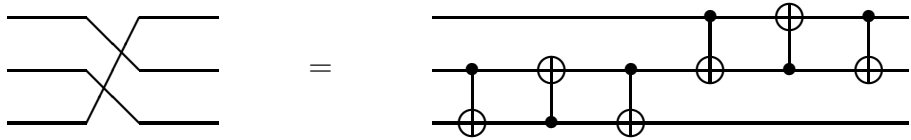}
\caption{Factorization $(1,3,2) = (1,2)(2,3)$}
\label{123}
\end{figure}

\item
Following the notation of \cite{Barenco:95} we denote a $k$-times 
controlled $U$ by $\Lambda_k(U)$. Lemma 7.2 and lemma 7.5 in 
\cite{Barenco:95} show that for $U\in{\cal U}(2)$ the gate 
$\Lambda_{n-1}(U)$ can be realized
with gate complexity $O(n^2)$ and $\Lambda_k(U)$ with $O(n)$ 
for $k < n-1$ on $n$ qubits.

\item
The Fourier transform $\DFT_{2^n}$ can be performed in $O(n^2)$
elementary operations on a quantum computer (see \cite{Shor:94},
\cite{Coppersmith:94}).

\item\label{shifter} Let $P_n \in S_{2^n}$ be the cyclic shift
permutation which acts on the states of the quantum register as $x
\mapsto x + 1 \; {\rm mod} \; 2^n$. Obviously the corresponding
permutation matrix is the $2^n$-cycle
\[ P_n = 
     \left(
\renewcommand{\arraystretch}{0.8}%
\setlength{\arraycolsep}{0.2em}%
        \begin{array}{ccccc}
           0 & 1 &   &  &\\
             & 0 & 1 &  & \\
             &   & \ddots & \ddots & \\
             &   &  & 0& 1 \\
           1 &   &  &  & 0 
        \end{array}
     \right).
\]
A factorization of $P_n$ as a product of multiple controlled NOTs
as shown in figure \ref{downshift} needs $O(n^2)$ basic operations%
\footnote{This quantum circuit, similar to the classical 
carry-look-ahead logic, has been found by Markus Grassl
following discussions with Amir Fijany in May 1998}.

\begin{figure}[hbt]
\input{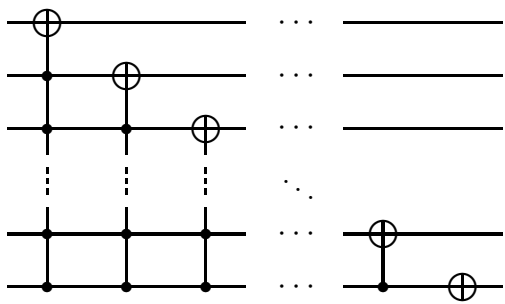}
\caption{\label{shift} Quantum circuit for the $2^n$-cycle $P_n$}
\label{downshift}
\end{figure}

\item
Let $U\in {\cal U}(2^n)$. The cost for a (single) controlled $U$ 
is settled by the following lemma.

\begin{lemma}\label{controledOP}
If $U\in {\cal U}(2^n)$ can be realized in $O(p(n))$ elementary
operations then $\Lambda_1(U)\in {\cal U}(2^{n+1})$ can also be
realized in $O(p(n))$ basic operations.
\end{lemma}

\noindent
Proof: First we assume without loss of generality that $U$ is
written in elementary gates. Therefore we have to show that a doubly
controlled NOT and a single controlled $U\in {\cal U}(2)$ can be
realized with a constant increase of length. This follows from
\cite{Barenco:95}.\qed
\end{itemize}

%
%

\section{Creating Fast Fourier Transforms}

In section \ref{GFT} we have explained that calculating a Fourier
transform for a group $G$ is the same as decomposing a regular
representation $\phi$ of $G$ with a matrix $A$ 
into irreducibles $\rho_i$ up to a
permutation $P$ such that equivalent irreducible summands are equal,
i.e.
$$
\phi^A = A^{-1}\cdot\phi\cdot A = 
\left(\rho_1\dirsum\dots\dirsum\rho_k\right)^P
\quad\mbox{fulfilling}\quad
\rho_i\cong\rho_j\Rightarrow\rho_i = \rho_j.
$$

A ``fast'' Fourier transform (on a classical computer) is given by a
factorization of $A$ into a product of sparse (w.r.t. the
architectural complexity measure) matrices.

In this section we present (without proof) a number of theorems and
lemmata yielding an algorithm to calculate fast Fourier transforms for
any solvable group $G$ (on a classical computer). The same algorithm
serves as a good starting point to obtain quantum Fourier transforms
if we assume the ``quantum wires'' to possess a suitable number of states.


The statements in this section all are taken from the first chapter of
\cite{Pueschel:98} where decomposition matrices and constructive
representation theory in general is investigated. There the objective
is the construction of decomposition matrices for monomial
representations which can be viewed as a generalization of Fourier
transforms.

The following theorem provides the crucial formula needed to obtain a
fast Fourier transform of $G$ by decomposing a regular representation
stepwise along a composition series of $G$. The formula has been known
(see \cite{Beth:84}) to yield fast Fourier transforms for solvable
groups on a classical computer (counting additions and
multiplications). Their tensor structure, however, also fits well to
the special architecture of a quantum computer. The general
constructive form of the following theorem as presented is due to
\cite{Pueschel:98} where furthermore an improved recursion formula
for the classical architecture can be found.

We use the following convention for the induction of a representation
$\phi$ of $H\leq G$ with transversal (i.e. a system of representatives for
the right cosets) $T = (t_1,\dots,t_n)$:
$$
(\phi\ind_T G)(x) = 
\left[\dot{\phi}(t_i x t_j^{-1})\mid i,j=1\dots n\right], 
$$
where $\dot{\phi}(y) = \phi(y)$ for $y\in H$ and the all-zero matrix else.
A regular representation $\phi$ is given by an induction 
$\phi = (1_E\ind_T G)$ where $1_E$ denotes the trivial 
representation of the trivial subgroup $E\leq G$.

\begin{figure}[ht]
\centerline{\unitlength2pt
\begin{picture}(65,40)(0,0)
\put(0,5){\vector(0,1){30}}
\put(0,0){\makebox(0,0){$N$}}
\put(0,40){\makebox(0,0){$G$}}
\put(25,5){\vector(0,1){30}}
\put(25,0){\makebox(0,0){$\phi$}}
\put(25,40){\makebox(0,0){$\overline{\phi}$}}
\put(24,20){\makebox(0,0)[r]{{\scriptsize$\rm ext$}}}
\put(30,0){\vector(1,0){30}}
\put(45,2){\makebox(0,0)[b]{\scriptsize$A$}}
\put(45,-2){\makebox(0,0)[t]{\scriptsize$\rm dec$}}
\put(65,5){\vector(0,1){30}}
\put(65,0){\makebox(0,0){$\rho$}}
\put(65,40){\makebox(0,0){$\overline{\rho}$}}
\put(64,20){\makebox(0,0)[r]{{\scriptsize$\rm ext$}}}
\end{picture}
}
\caption{Situation in theorem \ref{Induktionsrekursion}}
\label{fig: Induktionsrekursion}
\end{figure}
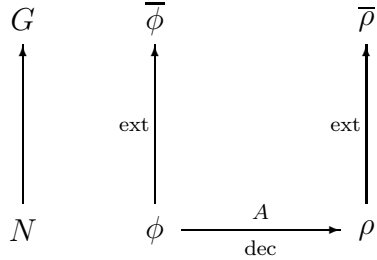

\begin{theorem}\label{Induktionsrekursion}
Let $N\normal G$ a normal subgroup of prime index $p$ with (cyclic)
transversal $T = (t^0,t^1,\dots,t^{(p-1)})$ and $\phi$ a
representation of degree $d$ of $N$ which has an extension
$\ext{\phi}$ to $G$ (see figure \ref{fig: Induktionsrekursion}).
Suppose that $A$ is matrix decomposing $\phi$ into irreducibles,
i.e. $\phi^A = \rho = \rho_1\dirsum\dots\dirsum\rho_k$ and that
$\ext{\rho}$ is an extension of $\rho$ to $G$.  Then
$$
B = (\onemat_p\tensor A)\cdot D\cdot(\DFT_p\tensor\onemat_d),
\quad\mbox{where}\quad
D = \bigdirsum_{i=0}^{p-1} \ext{\rho}(t)^i,
$$
is a decomposition matrix for $\phi\ind_T G$, more precisely
$$
(\phi\ind_T G)^B = \bigdirsum_{i=0}^{p-1}\lambda_i\cdot\ext{\rho},
$$
where $\lambda_i:\ t\mapsto\omega_p^i$, $i=0\dots p-1$, are 
the $p$ $1$-dimensional representations of $G$ arising from
the factor group $G/N$.
\end{theorem}

\noindent
In the case of an abelian group $G$ the formula yields exactly
the well-known Cooley-Tukey decomposition, hence $D$ can be
viewed as a generalized Twiddle factor. In the case of $G$ being a 
direct product $G\cong N\times G/N$ the Twiddle matrix $D$ vanishes.
Since we want to apply theorem \ref{Induktionsrekursion} to a regular
representation we need the following lemma.

\begin{lemma}\label{Erweiterung}
Let $N\normal G$ a normal subgroup of prime index $p$ and
$\phi$ any regular representation of $N$. Then $\phi$ 
(and hence all of its conjugates) has an extension $\ext{\phi}$ to $G$.
Furthermore $\phi\cong\phi^t$ for all $t\in G$.
($\phi^t:\ x\mapsto\phi(txt^{-1})$ is called the inner conjugate
of $\phi$ by $t$).
\end{lemma}

\noindent
In order to obtain an algorithm from theorem \ref{Induktionsrekursion}
we are faced with two problems. The first is the calculation
of the Twiddle matrix $D$ which is essentially the problem of 
extending $\rho$ to $\ext{\rho}$ and evaluating it at $t$. 
Suppose we are given $A$ and $\rho$ which is a direct sum of irreducibles, 
$\rho = \rho_1\dirsum\dots\dirsum\rho_k$, with equivalent summands 
being equal. 
Because of lemma \ref{Erweiterung}
$G/N$ operates on the irreducibles $\rho_i$ via 
inner conjugation (explained in the lemma above).
According to Clifford's Theorem (see e.g. \cite{Clausen:93}, pp.~88) 
exactly one of the following two cases applies to each 
summand $\rho_i$: Either $\rho_i \cong \rho_i^t$ 
and $\rho_i$ can be extended to $G$
or $\rho_i \not\cong \rho_i^t$ and $\rho_i\ind_T G$ is irreducible.
In the first case the extension may be calculated by 
Minkwitz' formula (see \cite{Minkwitz:96}), 
in the latter case the direct sum 
$\rho_i\dirsum\rho_i^t\dirsum\dots\dirsum\rho_i^{t^{(p-1)}}$ 
can be extended to $G$ by $\rho_i\ind_T G$.
We do not state Minkwitz' formula here since we will not need it
in the special cases treated in section \ref{2groups}.

The second problem arises from the fact that the 
decomposition $\bigdirsum_{i=0}^{p-1}\lambda_i\cdot\ext{\rho}$
in theorem \ref{Induktionsrekursion} does not satisfy the 
property of equivalent summands being equal. 
This can be achieved using the following lemma.

\begin{lemma}\label{Gleichmacher}
Let $N\normal G$ a normal subgroup of prime index $p$ with
transversal $T = (t^0,t^1,\dots,t^{(p-1)})$. 
Suppose that $\rho$ is an irreducible representation 
of degree $d$ of $N$ satisfying $\rho\not\cong\rho^t$ and 
$\lambda_i:\ t\mapsto\omega_p^i$ is 
an irreducible representation of $G$ arising from $G/N$.
Then 
$$
\left(\lambda_i\cdot(\rho\ind_T G)\right)^{D\tensor\onemat_d} = 
\rho\ind_T G,\quad D = \diag(1,\omega_p,\dots,\omega_p^{(p-1)})^i.
$$
\end{lemma}

\noindent
Now we are ready to formulate the algorithm which constructs
a fast Fourier transform for $G$ from a fast Fourier transform 
of a normal subgroup of prime index.

\begin{algorithm}\label{Algorithmus}
Let $N\normal G$ a normal subgroup of prime index $p$ with
transversal $T = (t^0,t^1,\dots,t^{(p-1)})$. Suppose that
$\phi$ is a regular representation of $N$ with decomposition
matrix $A$:
$$
\phi^A = \rho_1\dirsum\dots\dirsum\rho_k
\quad\mbox{fulfilling}\quad
\rho_i\cong\rho_j\Rightarrow\rho_i=\rho_j.
$$
A decomposition matrix $B$ for the regular representation
$\phi\ind_T G$ can be obtained as follows.
\begin{enumerate}
\item Determine a permutation matrix $P$ rearranging the $\rho_i$,
$i=1\dots k$, such that the extendable $\rho_i$ (i.e. those satisfying
$\rho_i=\rho_i^t$) come first followed by the others ordered into
sequences of length $p$ equivalent to 
$\rho_i,\rho_i^t,\dots,\rho_i^{t^{(p-1)}}$.
(Note: These sequences we need to equal 
$\rho_i,\rho_i^t,\dots,\rho_i^{t^{(p-1)}}$
which is established in the next step).
\item Calculate a matrix $M$ which is the identity on the 
extendables and conjugates the sequences of length $p$ to make
them equal to $\rho_i,\rho_i^t,\dots,\rho_i^{t^{(p-1)}}$.
\item Note that $A\cdot P\cdot M$ is a decomposition matrix 
for $\phi$, too, and let $\rho=\phi^{A\cdot P\cdot M}$.
Extend $\rho$ to $G$ summandwise. For the extendable summands
use Minkwitz' formula, the sequences 
$\rho_i,\rho_i^t,\dots,\rho_i^{t^{(p-1)}}$ can be extended by
$\rho_i\ind_T G$ as stated above.
\item Evaluate $\ext{\rho}$ at $t$ and build 
$\displaystyle D = \bigdirsum_{i=0}^{p-1}\,\ext{\rho}(t)^i$. 
\item Construct a blockdiagonal matrix $C$ with lemma \ref{Gleichmacher}
conjugating $\bigdirsum_{i=0}^{p-1}\lambda_i\cdot\ext{\rho}$
such that equivalent irreducibles are equal. $C$ is the identity 
on the extended summands.
\end{enumerate}
Then
\begin{eqnarray}\label{x}
B = (\onemat_p\tensor A\cdot P\cdot M)\cdot D\cdot
(\DFT_p\tensor\onemat_{|N|})\cdot C
\end{eqnarray}
is a decomposition matrix for $\phi\ind_T G$.\qed
\end{algorithm}
It is obviously possible to construct fast Fourier transforms on a
classical computer for any solvable group by recursive use of this
algorithm. Note that the consideration of T-adapted representations
(see \cite{Clausen:93}) here is unnecessary: The irreducibles are
constructed along with the decomposition matrices.

\begin{figure}
\input{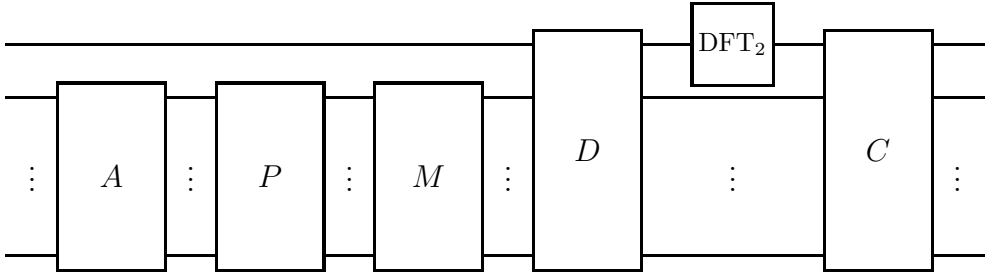}
\caption{\label{coarse} Coarse Quantum circuit visualizing 
algorithm \ref{Algorithmus}}
\end{figure}

Since we restrict ourselves to the case of a quantum computer
consisting of qubits, i.e. two-level systems, we apply algorithm
\ref{Algorithmus} to obtain QFTs for $2$-groups (size is a
$2$-power). In this case the two tensor products occuring in (\ref{x})
fit very well to yield a coarse factorization as shown in figure
\ref{coarse}. The remaining problem, however, is the realization of
the matrices $A,P,M,D,C$ in terms of elementary building blocks as
presented in section \ref{complexity}. At present this realization
remains a creative process which might be performed by hand if an
arbitrary class of groups is given. In section \ref{2groups} we will
apply algorithm \ref{Algorithmus} to a class of non-abelian 2-groups.

%
%

\section{Generating QFTs for a class of 2-groups}\label{2groups}

In the case of $G$ being an abelian $2$-group the realization of a
fast quantum Fourier transform has been settled by \cite{Kitaev:95}.
Clearly, this case is covered by the method presented here
(see the remarks following theorem \ref{Induktionsrekursion}).
In this section we will apply algorithm \ref{Algorithmus} to the class of
non-abelian $2$-groups containing a cyclic normal subgroup of index
2. Fast quantum Fourier transforms for these groups have already been
constructed by H\o yer in \cite{Hoyer:97}.

According to \cite{HuppertI:83}, p.~90/91 there are for $n\geq 3$
exactly four isomorphism types of non-abelian groups of order
$2^{n+1}$ affording a cyclic normal subgroup of order $2^n$:

\begin{itemize}
\item[(i)] The dihedral group 
$D_{2^{n+1}} = \langle x, y\mid x^{2^n} = y^2 = 1,\ x^y = x^{-1} \rangle$.

\item[(ii)] The quaternion group 
$Q_{2^{n+1}} = \langle x, y\mid x^{2^n} = y^4 = 1,\ x^y = x^{-1} \rangle$.

\item[(iii)] The group 
$QP_{2^{n+1}} = \langle x, y\mid x^{2^n} = y^2 = 1,\ 
x^y = x^{2^{n-1}+1} \rangle$.

\item[(iv)] The quasidihedral group 
$QD_{2^{n+1}} = \langle x, y\mid x^{2^n} = y^2 = 1,\ 
x^y = x^{2^{n-1}-1} \rangle$.
\end{itemize}

\noindent
Observe that the extensions (i), (iii), and (iv) of the cyclic subgroup
$Z_{2^n} = \langle x \rangle$ split, i.\,e. the groups have the
structure of a semidirect product of $Z_{2^n}$ by $Z_2$. 
The three isomorphism types correspond to the three different
embeddings of $Z_2 = \langle y\rangle$ into $(Z_{2^n})^\times
\cong Z_2 \times Z_{2^{n-2}}$.

%
%

\subsection{QFT for the dihedral groups $\bf D_{2^{n+1}}$} 

In this section we construct a QFT for the dihedral groups 
$D_{2^{n+1}}$ step by step according to algorithm \ref{Algorithmus}
and explicitly state the occuring quantum circuits.

Let $G = D_{2^{n+1}} = 
\langle x, y\mid x^{2^n} = y^2 = 1,\ x^y = x^{-1} \rangle$ with normal 
subgroup $N = \langle x\rangle \normal G$ of index 2 
and transversal $T = (1, y)$.
We consider the regular representation $\phi = (1_E\ind_S N)\ind_T G$ 
of $G$ with $S = (1,x,\dots,x^{2^n-1})$. Obviously the
regular representation $(1_E\ind_S N)$ of $N$ is decomposed 
by $A = \DFT_{2^n}$ into $\rho_0\dirsum\dots\dirsum\rho_{2^n-1}$
where $\rho_i = (x\mapsto \omega_{2^n}^i)$.
Now we are ready to apply algorithm \ref{Algorithmus} to obtain
a decomposition matrix $B$ for $\phi$.
For convenience we denote $\omega_{2^n}$ simply as $\omega$ 
and the Hadamard matrix as
$$
H = \DFT_2 = \frac{1}{\sqrt{2}} 
\left(
\begin{array}{rr} 1 & 1 \\ 1 & -1
\end{array}
\right).
$$
\begin{enumerate}
\item Since $\rho_i^y(x) = \rho_i(yxy^{-1}) = \rho_i(x^{-1}) 
= \rho_{2^n-i}(x)$ we see that there are exactly two extendable
$\rho_i$ namely for $i = 0, 2^{n-1}$. The sequences of inner conjugates
are given by $\rho_i, \rho_{2^n-i}$, $i \neq 0, 2^{n-1}$.
We need a permutation $P$ reordering the $\rho_i$ as
$$
\underbrace{\rho_0, \rho_{2^{n-1}}}_{\mbox{\scriptsize extendables}},\ 
\underbrace{\rho_1, \rho_{2^n-1},\dots,\rho_i,\rho_{2^n-i},\dots,
\rho_{2^{n-1}-1},\rho_{2^{n-1}+1}}_
{\mbox{\scriptsize pairs of inner conjugates}}.
$$
This can be accomplished by the circuit given in figure
\ref{dihedralperm} since the $n$-cycle on the qubits which is
performed first yields a decimation by two on the indices, i.\,e. the
indices $0, \ldots, 2^{n-1}-1$ have found their correct position. The
only thing which remains to do is to perform the operation $x\mapsto-x$ 
on the odd positions. This can be done by an inversion of all
(odd) bits followed by a $x\mapsto x+1$ shift $P_{n-1}$ 
on the odd states of the register.

\begin{figure}
\input{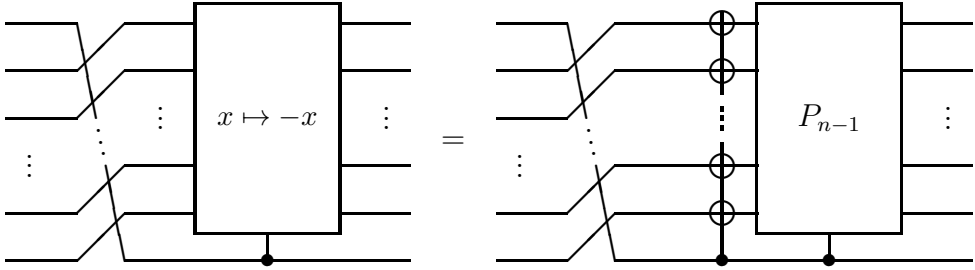}
\caption{\label{dihedralperm} Ordering the irreducibles of 
$Z_{2^n}\normal D_{2^{n+1}}$}
\end{figure}

\item $M$ can be omitted since all the $\rho_i$ are of degree 1.
\item Let $\phi^{A\cdot P} = \rho$. We extend $\rho$ summandwise
to $\ext{\rho}$:
\begin{itemize}
\item $\rho_0 = 1_N$ can be extended by $1_G$.
\item $\rho_{2^{n-1}}$ can be extended through 
$\ext{\rho}_{2^{n-1}}(y) = 1$.
\item The sequences $\rho_i\dirsum\rho_{2^n-i}$, $i \neq 0, 2^{n-1}$
can be extended by $\rho_i\ind_T G$:

\[ 
\rho_i\ind_T G : x\mapsto \left(
\begin{array}{cc}
\omega^i & 0 \\
0 & \omega^{-i}
\end{array}
\right),\ 
y \mapsto \left(
\begin{array}{cc}
0 & 1 \\
1 & 0 
\end{array}
\right).
\]

\end{itemize}
\item Evaluation of $\ext{\rho}$ at the transversal $T$ yields 
the Twiddle matrix
\begin{eqnarray*}
D & = & \ext{\rho}(1) \dirsum \ext{\rho}(y) \\
& = & \onemat_{2^n} \dirsum 
\renewcommand{\arraystretch}{0.2}%
\setlength{\arraycolsep}{0.2em}%
\left(\begin{array}{ccccccc}
1& & & & & \\
 &1& & & & \\
 & & &1& & & \\
 & &1& & & & \\
 & & & &\ddots& & \\
 & & & & & &1\\
 & & & & &1&
\end{array}\right).
\end{eqnarray*}
$D$ is realized by the quantum circuit given in figure 
\ref{dihedraltwiddle}. 

\begin{figure}[hbt]
\centerline{
\begin{minipage}[t]{0.47\linewidth}
\input{pic/twiddledihedral.pic}
\caption{\label{dihedraltwiddle} Twiddle matrix for $D_{2^{n+1}}$}
\end{minipage}
\qquad
\begin{minipage}[t]{0.47\linewidth}
\input{pic/cleanupdihedral.pic}
\caption{\label{dihedralclean} Equalizing inductions}
\end{minipage}}
\end{figure}

\item According to lemma \ref{Gleichmacher} the matrix $C$ has the
following diagonal form:
$$
C = \onemat_{2^n} \dirsum 
\diag(1,1,\underbrace{1,-1,\dots,1,-1}_
{2^{n-1}-1\; \mbox{\scriptsize pairs}}),
$$
which is realized by the quantum circuit given in figure \ref{dihedralclean}.
\end{enumerate}

\begin{figure}
\input{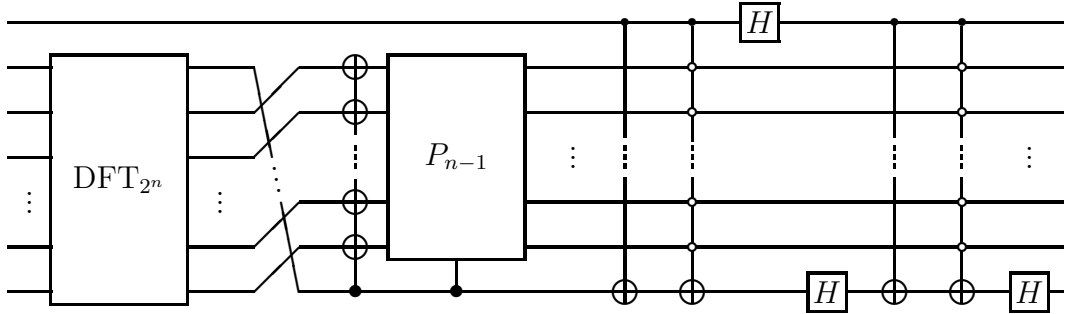}
\caption{\label{dihedralcomplete} Complete QFT circuit 
for the dihedral group $D_{2^{n+1}}$}
\end{figure}

\noindent
Summarizing we obtain that
$$
B = (\onemat_p\tensor A\cdot P\cdot M)\cdot D\cdot
(\DFT_p\tensor\onemat_{|N|})\cdot C
$$
is a decomposition matrix for $\phi$ and a fast quantum Fourier
transform for $G$. The whole circuit is shown in figure 
\ref{dihedralcomplete}.

%
%

\subsection{QFT for the groups 
$\bf Q_{2^{n+1}}$, $\bf QP_{2^{n+1}}$, and $\bf QD_{2^{n+1}}$}

In the following we give the circuits for the groups $Q_{2^{n+1}}$,
$QP_{2^{n+1}}$, and $QD_{2^{n+1}}$. In all cases we have $\langle
x\rangle = N\normal G$ so that algorithm \ref{Algorithmus} has to be
performed only once for the last step. For the sake of brevity we will
state only those parts of the circuit which differ from the dihedral
group. Some of the essential properties of the groups are summarized
in tabular \ref{twogroups}.  We use the same notation as in the last
section.

\begin{table}[hbt]
\renewcommand{\arraystretch}{1.1}
\centerline{
\begin{tabular}{|l|l|c|c|}
\hline
& & No. of 1-dim & No. of 2-dim\\
\raisebox{1.5ex}[-1.5ex]{Group} & 
\raisebox{1.5ex}[-1.5ex]{Inner conjugates of $Z_{2^n}$} 
& irreducibles & irreducibles \\
\hline
$D_{2^{n+1}}$  & $\rho_i,\quad \rho_{2^n-i}$ & $4$ & $2^{n-1}-1$\\
$Q_{2^{n+1}}$  & $\rho_i,\quad \rho_{2^n-i}$ & $4$ & $2^{n-1}-1$\\
$QP_{2^{n+1}}$ & $\rho_i,\quad \rho_{i(2^{n-1}+1)\,{\rm mod}\,2^n}$ & 
$2^n$ & $2^{n-2}$\\
$QD_{2^{n+1}}$ & $\rho_i,\quad \rho_{i(2^{n-1}-1)\,{\rm mod}\,2^n}$ & 
$4$ & $2^{n-1}-1$\\
\hline
\end{tabular}}
\renewcommand{\arraystretch}{1}
\caption{\label{twogroups} A class of non-abelian 2-groups}
\end{table}

\begin{figure}[hbt]
\input{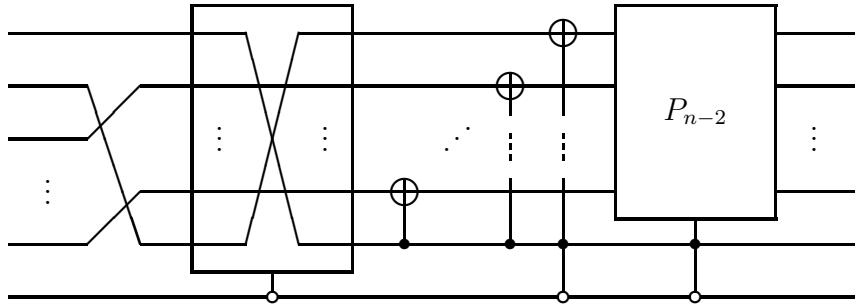}
\caption{\label{quasidihedralperm} The permutation for the $QD_{2^{n+1}}$}
\end{figure}

\begin{itemize}
\item $Q_{2^{n+1}}$: The irreducibles $\rho_i$ extend or induce 
in the same way as in the dihedral case. Hence the QFT only differs 
in the Twiddle matrix $D$ since for a not extendable $\rho_i$
we have 
$$
(\rho_i\ind_T G)(y) = 
\left(\!\!\!\begin{array}{rr}
0&1\\-1&0\end{array}\right).
$$
Thus the Twiddle matrix $D$ is given by
$$
D = \onemat_{2^n} \dirsum 
\renewcommand{\arraystretch}{0.2}%
\setlength{\arraycolsep}{0.2em}%
\left(\begin{array}{*{7}{c}}
1& & & & & \\
 &1& & & & \\
 & & &1& & & \\
 & &-1& & & & \\
 & & & &\ddots& & \\
 & & & & & &1\\
 & & & & &-1&
\end{array}\right)
$$
and can be realized by the circuit in figure \ref{quaterniontwiddle}.

\begin{figure}
\centerline{
\begin{minipage}[t]{0.47\linewidth}
\input{pic/twiddlequaternion.pic}
\caption{\label{quaterniontwiddle}Twiddle matrix for $Q_{2^{n+1}}$}
\end{minipage}
\qquad
\begin{minipage}[t]{0.47\linewidth}
\input{pic/qpgroup.pic}
\caption{\label{qpgroupperm} Permutation for $QP_{2^{n+1}}$}
\end{minipage}}
\end{figure}

\item $QP_{2^{n+1}}$: To determine which $\rho_i$ are extendable
we use $\rho_i^y(x) = \rho_i(yxy^{-1}) = \rho_i(x^{2^{n-1}+1})$.
Hence
$$
\rho_i = \rho_i^y\Leftrightarrow
\omega^i = \omega^{i\cdot(2^{n-1}+1)}\Leftrightarrow
\omega^{i\cdot 2^{n-1}} = 1\Leftrightarrow
2\mid i
$$
and there are exactly $2^{n-1}$ extendable $\rho_i$.
The reordering permutation $P$ has the easy form shown in figure
\ref{qpgroupperm}, and the matrix $D$ is given by
$$
D = \onemat_{2^n} \dirsum \onemat_{2^{n-1}}\dirsum
\renewcommand{\arraystretch}{0.2}%
\setlength{\arraycolsep}{0.2em}%
\left(\begin{array}{*{5}{c}}
&1& & & \\
1& & & & \\
& &\ddots& & \\
& & & &1\\
& & & 1&
\end{array}\right)
$$
which is simply a doubly controlled not as visualized in 
figure \ref{qptwiddle}.

The matrix $C$ then is given by figure \ref{qpequalizer}.

\begin{figure}
\centerline{
\begin{minipage}[t]{0.47\linewidth}
\input{pic/qptwiddle.pic}
\caption{\label{qptwiddle} Twiddle matrix for $QP_{2^{n+1}}$}
\end{minipage}
\qquad
\begin{minipage}[t]{0.47\linewidth}
\input{pic/qpequalizer.pic}
\caption{\label{qpequalizer} Equalizing for $QP_{2^{n+1}}$}
\end{minipage}}
\end{figure}

\item $QD_{2^{n+1}}$: Here we have $\rho_i^y(x) = \rho_i(x^{2^{n-1}-1})$
and
$$
\rho_i = \rho_i^y\Leftrightarrow
\omega^i = \omega^{i\cdot(2^{n-1}-1)}\Leftrightarrow
\omega^{i\cdot(2^{n-1}-2)}=1\Leftrightarrow
i=0, 2^{n-1}.
$$
Thus everything is the same as in the dihedral case beside the
ordering permutation $P$ which takes the more complicate form shown in
figure \ref{quasidihedralperm}.
\end{itemize}

\noindent
Investigation of the quantum circuits yields the following theorem.

\begin{theorem}
The Fourier transforms for the groups $G = D_{2^n},\ Q_{2^n},\
QP_{2^n}$, and $QD_{2^n}$ can be performed on a quantum computer in
$O(\log^2|G|)$ elementary operations.
\end{theorem}

\noindent
Proof: We can treat the four series uniformly, since the Fourier
transforms all have the same decomposition pattern. First, in all
cases a Fourier transform for the normal subgroup $Z_{2^{n-1}}$ is
performed with cost of $O(n^2)$ basic operations. The reordering
permutation $P$, the Twiddle matrix $D$, and the equalizing matrix $C$
cost $O(n^2)$ in case of $D_{2^n}$, $Q_{2^n}$, and $QD_{2^n}$ due to
lemma \ref{controledOP} and example \ref{shifter}. For $QP_{2^n}$ we
need only $O(1)$ operations for $P$, $D$, and $C$.\qed

\medskip
\noindent
All presented Fourier transforms have been implemented by the authors 
in the language GAP \cite{GAP} using the package AREP 
\cite{AREP} which will be available soon as a GAP share package.

%
%

\section{Conclusions and Outlook}

A constructive algorithm has been presented allowing to attack the
problem of constructing fast Fourier transforms for $2$-groups $G$ on
a quantum computer built up from qubits. For a certain class of
non-abelian $2$-groups the algorithm has been successfully
applied. All the QFTs created are of computational complexity
$O(\log^2|G|)$ like in the case of the cyclic group $Z_{2^n}$. The
main problem imposed by the implementation of certain permutation
and block diagonal matrices has been solved efficiently.

Using the recursion formula from theorem \ref{Induktionsrekursion} it
should be possible to construct QFTs for other classes of groups as
well as to realize certain signal transforms on a quantum computer by
means of symmetry-based decomposition (see \cite{Pueschel:98},
\cite{Egner:97}, \cite{Minkwitz:93}).

\medskip
\noindent
We are indebted to Markus Grassl for helpful comments and discussions.
Part of this work was presented and completed during the 1998
Elsag-Bailey -- I.S.I. Foundation research meeting on quantum
computation.

%
%

\bibliography{paper}
 
\bibliographystyle{plain}  

\end{document}